 \documentstyle[fleqn,11pt,openbib]{article}
 \let\huge=\large
 
 \let\large=\normalsize
 \newtheorem{Postulate}{Postulate}
 \newtheorem{Theorem}{Theorem}
 
\begin{document}
 \begin{flushleft}                   {\huge  \vspace*{10pc}
 REALIZATION OF MEASUREMENT               \\ \vspace*{4pt}
 AND THE STANDARD QUANTUM LIMIT           \\ \vspace*{2pc}
 \hspace*{1in}     Masanao Ozawa         }\\ \vspace*{1pc}
 \hspace*{1in}     Department of Mathematics            \\
 \hspace*{1in}     College of General Education         \\
 \hspace*{1in}     Nagoya University                    \\
 \hspace*{1in}     Nagoya 464, Japan
 \end{flushleft}
 \vspace*{2pc}
 %
 \section{Introduction}
 \label{1}
What measurement is there?  It is a difficult question but the
importance of this question has increased much in connection with
the effort to detect gravitational radiation.  For monitoring the
position of a free mass such as the gravitational-wave
interferometer \cite{W79}, it is usually supposed that the
sensitivity is limited by the so called standard
quantum limit (SQL) \cite{B74,C80}. In the recent controversy
\cite{Y83}--\cite{C85}, started with Yuen's proposal \cite{Y83} of a
measurement which beats the SQL, the meaning of the SQL has been
much clarified. In order to settle this controversy, rigorous
treatment of the question on what measurement there is seems the
key point.

Usually, the first approach to such an ontological question is
very mathematical.  Fortunately, in the last two decades,
mathematical theory for describing wide possibilities of quantum
measurement was developed in the field of mathematical physics
\cite{D76}--\cite{O86}. Indeed, the question on what measurement
there is was given a solution by the present author \cite{O84} under a
physically reasonable mathematical formulation. Unfortunately,
this result and the mathematics for this result have not been
familiar with physicists who need that result.

The purpose of this paper is two folds.  The first purpose is to
present the results from the mathematical theory of quantum
measurement in a form accessible for physicists.  As a
consequence of this theory, I shall settle the controversy of the
SQL in two ways; by a general consideration giving new vistas
concerning such a nonstandard measurement and by giving a model
of measuring interaction which breaks the SQL \cite{O88}.  The
second purpose is, of course, to present this result. 

 %
 \section{Standard Quantum Limit for Free-Mass Position}
 \label{2}

The uncertainty principle is a physicist's wisdom which gives
correct answers to many quantum mechanical problems without so much
cumbersome analysis of the problem. An application of this wisdom
to analysis of the performance of interferometric
gravitational-wave detector leads to the limit for sensitivity of
monitoring the free-mass position, which has long been a topic of
controversy within the quantum optics and general relativity
community.   This limit --- referred to as the {\em standard
quantum limit} (SQL) for monitoring the position of a free mass ---
is usually stated as follows: In the repeated measurement of its
position $x$ of a free mass $m$ with the time $\tau $ between two
measurements, the result of the second measurement cannot be
predicted with uncertainty smaller than $(\hbar \tau /m)^{1/2}$.

 \subsection{Yuen's proposal of breaching SQL}
 \label{2.1}

In the standard argument \cite{B74,C80} deriving this limit, the
uncertainty principle 
 \begin{equation}
 \Delta x(0)\Delta p(0) \geq \hbar /2                              \label{2.1.1}
 \end{equation}
 is applied to the position uncertainty $\Delta x(0)$ and the momentum
uncertainty $\Delta  p(0)$ at the time $t = 0$ just after the first
measurement so that by the time $\tau $ of the second measurement the
squared uncertainty (variance) of $x$ increases to
 \begin{equation}
 \Delta x(\tau )^2= \Delta x(0)^2 + \Delta p(0)^2\tau ^2/m^2 \geq  2\Delta x(0)\Delta p(0)\tau /m \geq  \hbar \tau /m.
                                                  \label{2.1.2}
 \end{equation}
 The SQL is thus obtained as
 \begin{equation}
 \Delta x(\tau ) \geq (\hbar \tau /m)^{1/2},                        \label{2.1.3}
 \end{equation}
 and it is usually explained as a straightforward consequence of
the uncertainty principle (\ref{2.1.1}).

Yuen \cite{Y83} pointed out a serious flaw in the standard
argument. Since the evolution of a free mass is given by 
 \begin{equation}
 \hat{x}(t) = \hat{x}(0) + \hat{p}(0)t/m,                        \label{2.1.4}
 \end{equation}
 the variance of $x$ at time $\tau $ is given by
 \begin{equation}
 \Delta x(\tau )^2 = \Delta x(0)^2 + \Delta p(0)^2\tau ^2/m^2
        + \langle \Delta \hat{x}(0)\Delta \hat{p}(0) + \Delta \hat{p}(0)\Delta \hat{x}(0)\rangle \tau /m   \label{2.1.5}
 \end{equation}
 where $\Delta \hat{x} = \hat{x} - \langle \hat{x}\rangle $ and $\Delta x^2 = \langle \Delta \hat{x}^2\rangle $, etc.  Thus the
standard argument implicitly assumes that the last term --- we
shall call it the {\em correlation term} --- in Eq.~(\ref{2.1.5})
is non-negative. Yuen's assertion \cite{Y83} is that some
measurements of $x$ leave the free mass in a state with the
negative correlation term.

 In probability theory, it is an elementary fact that the
variance of the sum of two random variables is the sum of their
variances plus the correlation term which is twice their
covariances. The covariance may be negative and it vanishes if
these random variables are independent.  In quantum mechanics, if
the state at $t = 0$ is a minimum-uncertainty one then the
correlation term vanishes. However, there are states with
negative correlation terms. The ratio of the covariance to the
product of the standard deviations is called the {\em
correlation coefficient} in probability theory. The negative
correlation expresses the tendency that the larger than the mean
one variable, the less than the mean the other. From this, one
can expect that in such a state the momentum works as
attracting the free mass around the mean position and
that the free evolution narrows the wave packet of the mass. 
Gaussian states \cite{S86} with this property are analyzed as
follows \cite{Y83}.

Let $a$ be the following operator in the Hilbert space of the
mass states:
 \begin{equation}
 \hat{a} = (m\omega /2\hbar )^{1/2}\hat{x} + 1/(2\hbar m\omega )^{1/2}\mbox{i}\hat{p},\quad [\hat{a},\hat{a}^{\dag}] = 1,
                                                  \label{2.1.6}
 \end{equation}
 where $\omega $ is an arbitrary parameter with unit sec$^{-1}$.  The
{\em twisted coherent state} (TCS) $|\mu \nu \alpha \omega \rangle $
is the eigenstate of $\mu \hat{a} + \nu \hat{a}^{\dag}$ with eigenvalue $\mu \alpha  + \nu \alpha ^*$:
 \begin{equation}
 (\mu \hat{a} + \nu \hat{a}^{\dag})|\mu \nu \alpha \omega \rangle  = (\mu \alpha  + \nu \alpha ^*)|\mu \nu \alpha \omega \rangle ,\quad
 |\mu |^2 - |\nu |^2 = 1.                             \label{2.1.7}
 \end{equation}
 The free-mass Hamiltonian is expressed by
 \begin{equation}
 \hat{H} = \hat{p}^2/2m = (\hbar \omega /2)(\hat{a}^{\dag}\hat{a} + \frac{1}{2} - \frac{1}{2}\hat{a}^2 -\frac{1}{2}\hat{a}^{\dag 2}).
                                                  \label{2.1.8}
 \end{equation}
 Within the choice of a constant phase the wave function
$\langle x|\mu \nu \alpha \omega \rangle $, where $\hat{x}|x\rangle  = x|x\rangle $, is given by
 \begin{equation}
 \langle x|\mu \nu \alpha \omega \rangle  = \left(\frac{m\omega }{\pi \hbar |\mu -\nu |^2}\right)^{1/4}
 \exp\left(-\frac{m\omega }{2\hbar }\frac{1+2\xi \mbox{i}}{|\mu -\nu |^2} (x-x_0)^2
 + \mbox{i}p_0(x-x_0)\right)                           \label{2.1.9}
 \end{equation}
where
 \begin{equation}
 \xi = \mbox{Im}(\mu ^*\nu ),\quad 
 \alpha = (m\omega /2\hbar )^{1/2}x_0 + 1/(2\hbar m\omega )^{1/2}\mbox{i}p_0, \quad
 x_0, p_0 \quad\mbox{real}.                       \label{2.1.10}
 \end{equation}

When $\xi = 0$, the wave function (\ref{2.1.9}) is the usual
minimum-uncertainty state.  In the context of oscillators, the
squeezed states are the wave functions of the form
(\ref{2.1.9})
with $\mu  \ne 0$.  The first two moments of $|\mu \nu \alpha \omega \rangle $ are
 \begin{eqnarray}
 \lefteqn{\langle x\rangle  = x_0, \quad \langle p\rangle  = p_0,}                      \label{2.1.11}\\
 \lefteqn{\Delta x^2 = \hbar |\mu - \nu |^2/2m\omega , \quad \Delta p^2 = \hbar m\omega |\mu + \nu |^2/2,}
                                                  \label{2.1.12}\\
 \lefteqn{\langle \Delta \hat{x}\Delta \hat{p} + \Delta \hat{p}\Delta \hat{x}\rangle  = -2\hbar \xi , \quad \Delta x\Delta p = \hbar (1+\xi ^2)^{1/2}/4,}
                                                  \label{2.1.13}\\
 \lefteqn{\langle H\rangle  = (\langle p\rangle ^2+ \Delta p^2)/2m.}                         \label{2.1.14}
 \end{eqnarray}
 When $\xi >0$ the $x$-dependent phase in (\ref{2.1.9}) leads to a
narrowing of the $\Delta x(t)$ from $\Delta x(0)$ during free evolution.
Because of this behavior, Yuen called mass states (\ref{2.1.9})
with $\xi >0$ as {\em contractive states}.

The position fluctuation for a free-mass starting in an arbitrary
TCS (\ref{2.1.9}) is immediately obtained from Eqs.~(\ref{2.1.5})
and (\ref{2.1.12})--(\ref{2.1.13}):
 \begin{eqnarray}
 \Delta x(t)^2&=&(\hbar /2m\omega )(|\mu -\nu |^2-4\xi \omega t + |\mu +\nu |^2(\omega t)^2)
                                                  \label{2.1.15}\\
         &=&(1/4\xi )(\hbar \tau /m) + (2\hbar /m\omega )(|\mu +\nu |\omega /2)^2(t-\tau )^2,
                                                  \label{2.1.16} 
 \end{eqnarray}
 where
 \begin{equation}
 \tau  =  2\xi /(\omega |\mu +\nu |^2) = \xi \hbar m/\Delta p(0)^2.           \label{2.1.17}
 \end{equation}
For $\xi > 0$, the position fluctuation decreases during time $t=0$
to $t=\tau $.  The minimum fluctuation achieves $1/(2\xi ^{1/2})$ times
the SQL
at the time $\tau = 2\xi /(\omega |\mu + \nu |^2)$.  Thus the minimum fluctuation
$\Delta x(\tau )$ is related only to the momentum uncertainty as follows:
 \begin{equation}
 \Delta x(\tau ) = \hbar /2\Delta p(0) =  \Delta x(0)/(1+4\xi ^2)^{1/2}.  \label{2.1.18}
 \end{equation}
This shows that $\Delta x(\tau )$ can be arbitrarily small for arbitrarily
large $\tau $ with sufficiently large $\xi $. It should be noted that
the minimum uncertainty product is realized between the momentum
uncertainty at $t=0$ and the position uncertainty at $t=\tau $.

Thus the SQL formulated by Eq.~(\ref{2.1.3}) can be breached if
there
is a measurement of free-mass position which leaves the free-mass
in a contractive state just after the measurement.

The reader may have the following question: Is the state after the
measurement always an eigenstate of the position observable? If so,
it never has any such contractive character.  It is natural to  say
so in the textbook description of measurement.  However, any real
measuring apparatus cannot leave the free mass in any eigenstate.
This statement has been repeatedly emphasized in the study of
measurement of continuous observables (observables with continuous
spectrum).  In the study of measurement, there is a deep gap
between discrete observables and continuous observables.  Once von
Neumann criticized Dirac's formulation in this point           
\cite[pp.222--223]{N55}: \lq \lq The division into quantized and
unquantized quantities corresponds $\ldots$ to the division into
quantities $R$ with an operator $\hat{R}$ that has a pure discrete
spectrum, and into such quantities for which this is not the
case.  And it was for the former, and only for these, that we
found a possibility of an absolutely precise measurement ---
while the latter could be observed only with arbitrarily good
(but never absolute) precision. (In addition, it should be
observed that the introduction of an eigenfunction which is
\lq improper,' i.e., which does not belong to Hilbert space $\ldots$
gives a less good approach to reality than our treatment here. 
For such a method pretends the existence of such states in which
quantities with continuous spectra take on certain values
exactly, although this never occurs. Although such idealizations
have often been advanced, we believe that  it is necessary to
discard them on these grounds, in addition to their mathematical
untenability.)''

 \subsection{Caves's defense of SQL}
 \label{2.2}

After Yuen's proposal \cite{Y83} of measurement with a contractive
state, Caves published a further analysis \cite{C85} of the SQL,
where he gave the following
improved formulation of the SQL:
 \begin{quotation}
 Let a free mass $m$ undergo unitary evolution during  the time
$\tau $ between two measurements of its position  $x$, made with
identical measuring apparatuses; the result of the second
measurement cannot be predicted with uncertainty smaller than
$(\hbar \tau /m)^{1/2}$, in average over the possible results of the
first measurement.
 \end{quotation}
 Caves \cite{C85} showed that the SQL holds for a specific model of
position measurement due to von Neumann \cite[pp.442--445]{N55}
and he also gave the following heuristic argument for the validity
of the SQL.  His point is the notion of the imperfect resolution
$\sigma $ of one's measuring apparatus.  His argument runs as follows.
{\em The first assumption} is that the variance $\Delta (\tau )^2$ of the
measurement of $x$ is the sum of $\sigma ^2$ and the variance of $x$ at
the time of the measurement; i.e.,
 \begin{equation}
 \Delta (\tau )^2 = \sigma ^2 + \Delta x(\tau )^2,                     \label{2.2.1}
 \end{equation}
and this is the case when  the measuring  apparatus is coupled
linearly
to $x$.
{\em The second assumption} is that just after the first
measurement, the free mass has position uncertainty not greater
than the resolution: 
 \begin{equation}
 \Delta x(0) \leq \sigma .                                     \label{2.2.2}
 \end{equation}
 Under these conditions, he derived 
the following estimate for the uncertainty $\Delta (\tau )$ of the second
measurement at time $\tau $:
 \begin{eqnarray}
 \Delta (\tau )^2&=&\sigma ^2+ \Delta x(\tau )^2\geq  \Delta x(0)^2+ \Delta x(\tau )^2\geq  2\Delta x(0)\Delta x(\tau )
                                                  \nonumber\\
         &\geq & \hbar \tau /m.                             \label{2.2.3}
 \end{eqnarray}
 According to this argument, the SQL is a consequence
from the uncertainty relation
 \begin{equation}
 \Delta x(0)\Delta x(\tau ) \geq  |\langle [\hat{x}(0),\hat{x}(\tau )]\rangle |/2 = \hbar \tau /2m, \label{2.2.4}
 \end{equation}
under assumptions (\ref{2.2.1})--(\ref{2.2.2}).

 However, his definition of the resolution of a measurement is
ambiguous and so a critical examination for his argument is
necessary. In fact, he used three different definitions in his
paper: 1) If the free mass is in a position eigenstate at the time
of a measurement of $x$, then $\sigma $ is defined to be the uncertainty
in the result. 2) The measurement determines the position
immediately after the measurement to be within roughly a distance
$\sigma $ of the measured value. 3) The square $\sigma ^2$ of the resolution
is the variance of the pointer-position just before the
system-meter interaction.  These three notions are essentially
different, although they are the same for von Neumann's model.

The notion of resolution of measurement has been often mentioned in
literature but up to now we have not yet reached any satisfactory
definition for it. What does the readout value tell one the states
of the free mass? There are two principal thoughts about this
question. One thinks that the readout tells the position of the
free mass just before the measurement, since the prior position is
the cause of the effect of the measurement that is the readout
value. Another one thinks that the readout tells the position of
the free mass just after the measurement, since the measurement
changes the position of the free mass so that the two effects of
the measurement --- the posterior position and the readout --- may
be in concordance.  Which is true is hardly answered. The best way
to attack the problem seems to recognize that there are two types
of concept of resolution of measurement. From this reason, we make
the following distinction: If the free mass is in a position
eigenstate at the time of a measurement of $x$ then the {\em
precision} $\varepsilon $ of the measurement is defined to be the uncertainty
in the result and the {\em resolution} $\sigma $ is defined to be the
deviation of the position of the free mass just after the
measurement from the readout just obtained. The mathematical
definitions of these concepts including the case of superposition
will be given and discussed thoroughly in the later sections.

We shall return to the problem of the validity of the SQL. By the
improvement of the formulation of the SQL, the idea of measurement
leaving the free mass in a contractive state can not readily
vitiate the SQL. However, there is a possibility for circumventing
the heuristic proof given by Caves \cite{C85}, since his assumption
$\sigma \geq \Delta x(0)$ is formulated in an ambiguous manner.  Nonetheless,
it is a heavy burden for one who wants to vitiate the SQL to make
a realizable model of measurement which circumvents Caves's
assumption. In the next section, we shall give general
considerations of realization of measurement.

Before going further, I shall mention certain attempts of
breaking the SQL. Recently, Ni \cite{N86} proposed a scheme of 
repeated position measurements for which one can predict the
result of the next measurement with arbitrarily small errors.
However, a close examination of this scheme leads to the
conclusion that the improved formulation of the SQL due to Caves
\cite{C85} is not broken by this repeated measurement scheme.
This scheme uses a combination of the Arthurs-Kelley measurements
\cite{A65}  which measure the position and momentum
simultaneously and approximately. Accordingly, one measurement of
this scheme has four meters, two of which measure the free-mass
position with a high resolution by one meter and with a low
resolution by the other meter, and the other two of which measure
the free-mass momentum with lower resolutions. The prediction of
the result of the next position measurement is done using these
four readouts.  If one of these meter is left alone, the
prediction cannot have the desired accuracy. This means that one
meter reading with the highest position resolution serves the
position measurement and the other three meter readings serve the
preparation procedure for the next measurement. There are several
similar models proposed with preparation
procedures for the next measurement and these examples do not
vitiate the improved SQL, since it disallows explicitly any
tinkering between two identical position measurements.  Indeed,
in these proposed models (e.g., \cite{N86,YU}), there is
at least one auxiliary meter with a lower resolution, the reading
of  which prepares the state for the next position measurement
really done by the other high resolution meter-reading. Thus any
many-meter systems are excluded out of the scope of the improved SQL.
The true problem is thus whether we can make a measurement with
only one meter, the reading of which gives the precise position
of the mass and simultaneously prepares the state for the next
identical position measurement, for instance, in a contractive
state.

 %
 \section{Quantum Mechanics of Measurement}
 \label{3}

What can one say about quantum measurement from quantum mechanics?
Von Neumann is acknowledged to be the first scientist who made a
route to this problem. Although his original motivation was to show
the consistency of the repeatability hypothesis (usually referred
to as the projection postulate) with the axioms of quantum
mechanics, his argument opened the way to analyze the quantum
measurement within quantum mechanics. However, his method caused
the controversy about the difficult problem of interpretation of
quantum mechanics. In what follows, I shall attempt to present von
Neumann's method with some elaborations which avoid difficulties of
the problem of interpretation and discuss several consequences from
the quantum mechanics of measurement.

 \subsection{Statistics of measurement}
 \label{3.1}

By the axioms of quantum mechanics we shall mean the axioms of
nonrelativistic quantum mechanics without any superselection
rules, which are reduced to (a) the definitions of observables
and states as self-adjoint operators and state vectors of a
Hilbert space, (b) the Schr\"{o}dinger time-dependent equation and
(c) the Born statistical formula for probability distributions of
commuting observables.  The projection postulate is excluded from
our axioms and its status will be discussed below. 

In order to discuss all possible quantum measurements, it is
convenient to classify them into two classes.  A measurement is of
the {\em first kind} if it does not destroy the quantum mechanical
description of the system to be measured so that we can determine
in principle the state just after the measurement corresponding to
the result of measurement.  A measurement is of the {\em second
kind} if it destroys the quantum mechanical description of the
system.  The whole process of a direct interaction with a
macroscopic detector such as a photon counter is considered as a
measurement of the second kind.

The statistics of a measurement of the first kind is specified by
the following two elements: For the system state $\psi $ just before
the measurement, let $P(a|\psi )$ the probability density of obtaining
the result $a$ and let $\psi _a$ be the system state just after the
measurement with result $a$. Then the physical design and the
indicated preparation of the apparatus determine $P(a|\psi )$ and the
transition $\psi \to  \psi _a$ for all possible $\psi $. These two elements will
be called the {\em statistics} of a given measurement of the first
kind; $P(a|\psi )$ will be called the {\em measurement probability}
and $\psi \to  \psi _a$ will be called the {\em state reduction}, further we
shall call the state $\psi $ just before the measurement as the {\em
prior state} and the state  $\psi _a$ just after the measurement as the
{\em posterior state}. (For notational convenience, $\psi _a$ will be
denoted sometimes by $\psi [a]$.)  This specification of measurement
statistics implies that if two given measurements are identical
then  the corresponding two statistics are identical.  On the
other hand, the statistics of a measurement of the second kind is
specified only by its measurement probability, for such a
measurement does not allow to describe the system state after the
measurement.

 \subsection{Scheme of measurement}
 \label{3.2}

Once we accept the axioms of quantum mechanics, it is natural
to accept in principle the following fact as a basis of
our discussion.

 \begin{Postulate}\label{P1}  
 For any observable $A$ with its spectral decomposition 
 \begin{equation}
 \hat{A} = \int x\,d\hat{A}(x),                    \label{3.2.1}
 \end{equation}
there is a measurement which may be of the second kind such that
its measurement probability $P(a|\psi )$ satisfies the Born
statistical formula
 \begin{equation}
 P(a|\psi )da = \langle \psi |d\hat{A}(a)|\psi \rangle ,                      \label{3.2.2}
 \end{equation}
for all prior state $\psi $.
 \end{Postulate}

We shall call any measuring apparatus satisfying Eq.(\ref{3.2.2})
as a {\em detector} for an observable $A$.
It should be noticed that Postulate~\ref{P1} alone never implies
existence of any measurements of the first kind.  Our fundamental
problem is thus {\em what measurement of the first kind is allowed
within our axioms and postulates of quantum mechanics}.  In order
to solve this problem, we adopt the following scheme of measurement
instituted by von Neumann.

Suppose that a quantum system $S$ --- called the {\em object
system} --- with the unknown system state $\psi $ just before the
measurement is to be measured by a measuring apparatus.  For the
observer --- called the {\em first observer} --- who applies
quantum mechanics only to the object system, this measurement is
described by the statistics of this measurement specified by the
given measuring apparatus.  Suppose that this measurement is of the
first kind and we shall denote the statistics of this measurement
by $P_{I}(a|\psi )$ and $\psi \to  \psi _a$. Then by nondestructive nature of
measurement of the first kind, there are other possibilities of
application of quantum mechanics to this physical phenomena of the
measurement. One possibility of such a quantum mechanical
description of measurement arises if one separates the measuring
apparatus into two parts. The first part --- called the {\em probe
system} --- is a microscopic system which directly interacts with
the object system and the second part is a detector which makes a
second kind measurement of an observable --- usually called the
pointer position in somewhat misunderstanding manner --- of the
probe system. 

For the observer --- called the {\em second observer} --- who
applies quantum mechanics to the composite system of the object and
the probe, this measurement is described as a combination of an
object-probe interaction and a second kind measurement of the probe
in the following manner.  Let $P$ be the probe system.  By the
arrangement of the measuring apparatus the following elements can
be specified as controllable elements; the system state $\varphi $ of the
probe system just before the measurement, the time evolution $\hat{U}$ of
the composite system $S+P$ during the interaction and the
observable $A$ of the probe system to be measured by the detector. 
Then just before the interaction the state of the composite system
is $\psi \otimes \varphi $ and hence just after the interaction the composite
system is in the state $\hat{U}(\psi \otimes \varphi )$. What can the second observer
tell about the statistics of this measurement? The measurement
probability $P_{II}(a|\psi )$ for the second observer is obviously the
postulated probability distribution of the observable $A$ in the
state $\hat{U}(\psi \otimes \varphi )$:
 \begin{equation}
 P_{II}(a|\psi )\,da = \langle \hat{U}(\psi \otimes \varphi )|1\otimes d\hat{A}(a)|\hat{U}(\psi \otimes \varphi )\rangle  .
                                                  \label{3.2.3}
 \end{equation}
Thus from the consistency of the measurement probabilities of these
two observers, we have
 \begin{equation}
 P_{I}(a|\psi )\,da = \langle \hat{U}(\psi \otimes \varphi )|1\otimes d\hat{A}(a)|\hat{U}(\psi \otimes \varphi )\rangle  .
                                                  \label{3.2.4}
 \end{equation}

In order to determine the state reduction, we may assume that
immediately after the interaction the object system would be
subjected to a detector of an arbitrary observable $X$ of the
object system. Since the probe system is also to be subjected to
the detection of $A$ immediately after the interaction in the
second-observer description, quantum mechanics predicts the joint
probability density $P_{II}(x,a|\psi )$ of obtaining
the result $A = a$ and $X = x$ for the second observer by the
relation
 \begin{equation}
 P_{II}(x,a|\psi )\,da\,dx = \langle \hat{U}(\psi \otimes \varphi )|d\hat{X}(x)\otimes d\hat{A}(a)|\hat{U}(\psi \otimes \varphi )\rangle  .
                                                  \label{3.2.5}
 \end{equation}
 For the first observer, the detection of the observable $X$ occurs
in the system state $\psi _a$ given the result $a$ of the first
measurement and hence the probability density $P_{I}(x,a|\psi )$ of the
same event is calculated by the first observer as follows:
 \begin{equation}
   P_{I}(x,a|\psi )\,da\,dx = \langle \psi _a|d\hat{X}(x)|\psi _a\rangle P_{I}(a|\psi )\,da.
                                                  \label{3.2.6}
 \end{equation}
 From the consistency of the statistics of these two observers, we
have
 \begin{equation}
  \langle \psi _a|d\hat{X}(x)|\psi _a\rangle  =
 \frac{\langle \hat{U}(\psi \otimes \varphi )|d\hat{X}(x)\otimes d\hat{A}(a)|\hat{U}(\psi \otimes \varphi )\rangle }
         {\langle \hat{U}(\psi \otimes \varphi )|1\otimes d\hat{A}(a)|\hat{U}(\psi \otimes \varphi )\rangle }.     \label{3.2.7}
 \end{equation}
The arbitrariness of $X$ yields the following relation for any
basis $\{|i\rangle \}$,
 \begin{equation}
 \langle i|\psi _a\rangle \langle \psi _a|j\rangle  = \frac{\langle \hat{U}(\psi \otimes \varphi )|\left(|j\rangle \langle i|\otimes d\hat{A}(a)\right)|\hat{U}(\psi \otimes \varphi )\rangle }
                              {\langle \hat{U}(\psi \otimes \varphi )|1\otimes d\hat{A}(a)|\hat{U}(\psi \otimes \varphi )\rangle },
                                                  \label{3.2.8}
 \end{equation}
and consequently,
 \begin{equation}
 |\psi _a\rangle \langle \psi _a| 
 = \sum _{i,j} |i\rangle  \frac{\langle \hat{U}(\psi \otimes \varphi )|\left(|j\rangle \langle i|\otimes d\hat{A}(a)\right)|\hat{U}(\psi \otimes \varphi )\rangle }
                          {\langle \hat{U}(\psi \otimes \varphi )|1\otimes dA(a)|\hat{U}(\psi \otimes \varphi )\rangle }      \langle j|.
                                                  \label{3.2.9}
 \end{equation}

Thus we have shown that the second observer can calculate the
statistics of this measurement from his knowledge about the
measuring apparatus and predict what will
happen to the first observer.  For example, if the first observer
observes that the system state $\psi $ is reduced to the state $\psi _a$
just after the measurement with the readout value $a$, the second
observer can calculate just the same state reduction $\psi  \to  \psi _a$ by
Eq.(\ref{3.2.9}).  The difference between those two observers is
that the first observer knows only the relation between the readout
$a$ and the reduced state $\psi _a$ but the second observer does know
the statistical correlation between the object system and the probe
system from which he can predict all statistics of this
measurement.  Some authors have made a misunderstanding at this
point.  They usually say that the state reduction is a consequence
of the observation or of the knowing of the readout.  If this
would be the case, we would get an obvious contradiction between
the first and the second observer; in fact, then the first observer
would say that the state reduction occurs after the detection of
the probe system contrary to the second observer's saying that it
occurs already before the detection of the probe system just after
the object-probe interaction.  The point is that there is no
causality relation between the readout $a$ and the reduced state
$\psi _a$ but
there is only a statistical correlation. Usually, statistical
correlation between two events does not imply any causality
relation.  In fact, for another observer who observes the result of
succeeding  measurement of the object system first, the state
reduction of the prove system occurs. We can only say that there is
a statistical correlation between the results of successive
measurements. Quantum mechanics shows that to get information from
one system is to make a statistical correlation with another system
by an interaction to be described by quantum mechanics. However,
there is another problem --- indeed, a different problem.  When
does the
statistical interference between the object system and the probe
system vanish?  This is the content of Schr\"{o}dinger's cat
paradox. This is a deep problem. However, in our formalism, we can
avoid the difficulty --- as promised before --- by boldly saying
\lq \lq During the second kind measurement of the probe system.''

 \subsection{Operation measures and measurement statistics}
 \label{3.3}

From the preceding analysis of the process of quantum measurement,
the problem as to what measurement there is can be reduced to the
problem what statistics can be realized by the measurement scheme. 
An important step to the mathematical solution of the latter
problem is to get a neat mathematical representation of the
statistics of measurement.  In what follows, we shall show that any
plausible description of measurement statistics can be expressed by
a single mathematical object called an operation measure.  A
thorough discussion of operations and effects may be found in
\cite{L83,K83}, and of operation measures and effect
measures in \cite{D76,H82,O84,O85}.

Let ${\cal H}$ be a Hilbert space of an object system.  Suppose that a
statistics of a measurement is given; this means that for any state
vector $\psi $ in ${\cal H}$ the measurement probability $P(a|\psi )$ and the
state reduction $\psi  \to  \psi _a$ is presupposed. Our first task is to
extend this statistical description to the case that the prior
state is a mixture. Let $\hat{\rho}$ be a density operator which represents
the prior state with its spectral decomposition
 \begin{equation}
     \hat{\rho} = \sum _i \lambda _i|\psi ^i\rangle \langle \psi ^i|.                 \label{3.3.1}
 \end{equation}
In this case, the measurement probability --- denoted by $P(a|\hat{\rho} )$
--- is given by
 \begin{equation}
 P(a|\hat{\rho} ) = \sum _i \lambda _i P(a|\psi ^i).                   \label{3.3.2}
 \end{equation}
Then the posterior state $\hat{\rho}_a$ of the system for the readout $a$
is obviously a mixture of all $|\psi ^i_a\rangle \langle \psi ^i_a|$'s , where $\psi ^i_a$
is the posterior state for the prior state $\psi ^i$, such that their
relative frequency is proportional to $\lambda _i$ and $P(a|\psi ^i)$.  Thus
we have
 \begin{equation}
 \hat{\rho}_a = \frac{\sum _i\lambda _iP(a|\psi ^i)|\psi ^i_a\rangle \langle \psi ^i_a|}
              {\sum _i\lambda _iP(a|\psi ^i)}.                \label{3.3.3}
 \end{equation}

Let $\Delta $ be an interval --- or more generally a Borel set --- in
the real line and let $S_\Delta $ the subensemble of the object system
selected by the condition that the readout $a$ of this measurement
lies in $\Delta $. Then we can ask what is the state of $S_\Delta $--- denoted
by $\hat{\rho}_\Delta $ --- just after the measurement. It is well known that the
state of this ensemble is not the superposition of all $\psi _a$'s with
$a$ in $\Delta $ but a mixture of all $|\psi _a\rangle \langle \psi _a|$'s with $a$ in $\Delta $;
their relative frequency is proportional to $P(a|\psi )$.  Thus the
state $\hat{\rho}_\Delta $ is
 \begin{equation}
 \hat{\rho}_\Delta  = \frac{\int _\Delta  da\,P(a|\psi )|\psi _a\rangle \langle \psi _a|}   
              {\int _\Delta  da\,P(a|\psi )}.                 \label{3.3.4}
 \end{equation}
 Further, for the case that the prior state is given by a mixture
represented by a density operator (\ref{3.3.1}), the state $\hat{\rho}_\Delta $
of the ensemble $S_\Delta $ just after the measurement is
 \begin{equation}
 \hat{\rho}_\Delta  = \frac{\int _\Delta  da\,P(a|\hat{\rho} )\hat{\rho}_a}
              {\int _\Delta  da\,P(a|\hat{\rho})},                \label{3.3.5}
 \end{equation}
where $\hat{\rho}_a$ is given in Eq.(\ref{3.3.3}).

These considerations lead to the following mathematical definition
of the transformation ${\cal I}(\Delta )$ which maps a density operator $\hat{\rho}$
to a trace class operator ${\cal I}(\Delta )\hat{\rho}$ given by the relations
 \begin{equation}
     {\cal I}(\Delta )\hat{\rho} = \int _\Delta  d{\cal I}(a)\hat{\rho}, \quad
     d{\cal I}(a)\hat{\rho} = \hat{\rho}_aP(a|\hat{\rho})\,da.                 \label{3.3.6}
 \end{equation}
An obvious requirement for this transformation ${\cal I}(\Delta )$ is as
follows:  If $\hat{\rho}$ is the mixture of $\hat{\sigma}_1$ and $\hat{\sigma}_2$ then ${\cal I}(\Delta )\hat{\rho}$
is the mixture of ${\cal I}(\Delta )\hat{\sigma}_1$ and ${\cal I}(\Delta )\hat{\sigma}_2$.   From this ${\cal I}(\Delta ) $
can be extended to the mapping of all trace class operators.  Then
the following properties can be easily deduced from
Eq.(\ref{3.3.6}):
 \begin{enumerate}
 \item For any Borel set $\Delta $, the mapping $\hat{\rho} \to   {\cal I}(\Delta )\hat{\rho}$
is a linear transformation on the space of trace class operators
which maps density operators to a positive trace class operator.
 \item For any Borel set $\Delta $ and any density operator $\hat{\rho}$,
 \begin{equation}
 0 \leq \mbox{Tr}[{\cal I}(\Delta )\hat{\rho})] \leq 1 \quad \mbox{and}\quad
 \mbox{Tr}[{\cal I}({\bf R})\hat{\rho}] = 1.                        \label{3.3.7}
 \end{equation}
 \item For any countable disjoint sequence $\{\Delta _i\}$ of Borel sets
and density operator $\hat{\rho}$,
 \begin{equation}
 \sum _i\mbox{Tr}[{\cal I}(\Delta _i)\hat{\rho}] = \mbox{Tr}[{\cal I}(\bigcup _i\Delta _i)\hat{\rho}].
                                                  \label{3.3.8}
 \end{equation}
 \end{enumerate}
A mapping $\hat{\rho} \to  {\cal I}(\Delta )\hat{\rho}$ with properties 1--2 is called an {\em
operation} and a mapping $\Delta  \to  {\cal I}(\Delta )$ with all properties 1--3 is
called an {\em operation measure}. Thus we have shown that if we
are given a plausible statistical description of measurement
we can construct an operation measure. The mathematical importance
of the operation measures is that it unifies two statistical data
of measurement --- the measurement probability and the state
reduction --- into a single mathematical object. In fact, the
measurement probability can be recovered by the relation
 \begin{equation}
 P(a|\hat{\rho})da = \mbox{Tr}[d{\cal I}(a)\hat{\rho}],               \label{3.3.9}
 \end{equation}
and the state reduction can be retained by the relation
 \begin{equation}
 \hat{\rho}_a = \frac{d{\cal I}(a)\hat{\rho}}{\mbox{Tr}[d{\cal I} (a)\,\hat{\rho}]}.
                                                  \label{3.3.10}
 \end{equation}
The mathematical justification of the above differentiation is
given in \cite{O85} for arbitrary operation measure --- so that,
without any presupposition of measurement statistics,
any operation measure gives a measurement statistics.
To put it simply, integration of any measurement statistics
gives an operation measure and differentiation of any operation
measure gives a measurement statistics.

Let $F$ be a (random) variable the values of which shows the
readout of a measurement. Then the probability distribution
$P(F\in \Delta |\hat{\rho})$ that the value of $F$ is in $\Delta $ given the prior
state $\hat{\rho}$ is obtained from Eq.(\ref{3.3.9}), i.e., 
 \begin{equation}
 P(F\in \Delta|\hat{\rho}) = \int _\Delta P(a|\hat{\rho})da.
 \end{equation}
In this case there is a unique positive operator
valued measure $\Delta \to \hat{F}(\Delta )$ such that
 \begin{eqnarray}
 \lefteqn{0 \leq \hat{F}(\Delta ) \leq 1, \quad\mbox{and}\quad \hat{F}({\bf R})=1,}  
                                             \label{3.3.11}\\
 \lefteqn{\mbox{Tr}[\hat{F}(\Delta )\hat{\rho}] = P(F\in \Delta |\hat{\rho}).}            
                                                  \label{3.3.12}
 \end{eqnarray}
We shall call this $\hat{F}$ as the {\em effect measure} of an
operation measure ${\cal I}$. If an operation measure ${\cal I}$ represents an
exact measurement of an observable $\hat{X} = \int x\,d\hat{X}(x)$,
then the corresponding effect measure $\hat{F}$ satisfies
 \begin{eqnarray}
 P(a|\hat{\rho})da &=& \mbox{Tr}[d\hat{X}(a)\hat{\rho}], \nonumber\\
 \mbox{Tr}[\hat{F}(\Delta )\hat{\rho}] &=& \int _\Delta  \mbox{Tr}[d\hat{X}(x)\hat{\rho}],     \label{3.3.13}
 \end{eqnarray}
and hence the effect measure of ${\cal I}$ coincides with the spectral
measure of operator $\hat{X}$, i.e., $d\hat{F}(x) = d\hat{X}(x)$.
In this sense, the notion of an effect measure generalizes
the conventional presupposition that the measurement probability
is represented by a spectral measure to the case of non-exact
measurements. 

 \subsection{Characterization of realizable measurement}
 \label{3.4}

In the preceding subsections, we have shown the following two
facts: 
 \begin{enumerate}
 \item  Any measurement scheme consisting of an object-probe
interaction and a probe detection determines the unique
measurement statistics by means of the second-observer
description.
 \item Any plausible measurement statistics of the first-observer
description gives an operation measure which unifies the
measurement statistics in a single mathematical object.
 \end{enumerate}
However, it is not at all clear whether every operation
measure is consistent with the second-observer description of
the measurement --- or what operation measures are consistent
with the second-observer description.

This problem has the following rigorous mathematical formulation:
Let ${\cal H}$ be a Hilbert space describing the object system and let
${\cal I}$ be an operation measure for the space ${\cal T}({\cal H})$ of the trace
class operators on ${\cal H}$.  The problem is to determine when we can
find another Hilbert space ${\cal K}$ describing the probe system with
a self-adjoint operator $\hat{A}$ in ${\cal K}$ describing the observable
actually detected, a state vector $\varphi $ in ${\cal K}$ describing the
preparation of the probe system and a unitary operator $\hat{U}$ on
${\cal H}\otimes {\cal K}$ describing time evolution of the object-probe composite
system during the measurement interaction such that this
second-observer description of measurement leads to the same
measurement statistics as the first-observer description of the
given operation measure ${\cal I}$.  For the last part of this
formulation, recall that the second-observer description of the
measurement leads to the statistics given by Eqs.(\ref{3.2.3}) and
(\ref{3.2.9}). On
the other hand, the operation measure is given by
Eq.(\ref{3.3.6}).
Thus
the condition for these two to give the same statistics is the
following relation:
 \begin{equation}
 \langle i|d{\cal I}(a)(|\psi \rangle \langle \psi |)|j\rangle  =
\langle \hat{U}(\psi \otimes \varphi )|\left(|j\rangle \langle i|\otimes d\hat{A}(a)\right)|\hat{U}(\psi \otimes \varphi )\rangle   \label{341}
 \end{equation}
for all $\psi $ and a basis $\{|i\rangle \}$ in ${\cal H}$.
We shall call any operation measure satisfying Eq.(\ref{341}) for some 
$\hat{A}$, $\phi$ and $\hat{U}$ as {\em realizable}.

In order to
present the solution of the problem which has been obtained in
\cite{O84}, we need one more mathematical concept
concerning the positivity property of the operation.
Let ${\cal I}$ be an operation measure.  Then the transformation
$\hat{\rho} \to  {\cal I}(\Delta )\hat{\rho}$ on the trace class operators is positive,
in the sense that for any density operator $\hat{\rho}$ the trace class operator
${\cal I}(\Delta )\hat{\rho}$ is a positive operator.  It follows that for any
vector $\varphi $, $\psi $ we have
 \begin{equation}
 \langle \varphi |d{\cal I}(a)(|\psi \rangle \langle \psi |)|\varphi \rangle  \geq 0.                    \label{3411}
 \end{equation}
An operation measure is called {\em completely positive} if it has
the
following stronger positivity property; for any finite
sequences of vectors $\xi _1,\xi _2,$\ldots$,\xi _n$ and $\eta _1,\eta _2,$\ldots$,\eta _n$,
 \begin{equation}
 \sum _{i,j=1}^{n} \langle \xi _i|d{\cal I}(a)(|\eta _i\rangle \langle \eta _j|)|\xi _j\rangle  \geq 0.
                                                  \label{3412}
 \end{equation}
If an operation measure is realizable then by Eq.(\ref{341})
we
obtain 
 \begin{eqnarray*}
 \lefteqn{\sum _{i,j=1}^{n} \langle \xi _i|d{\cal I}(a)(|\eta _i\rangle \langle \eta _j|)|\xi _j\rangle }\\
 & & = \sum _{i,j=1}^{n}
  \langle \hat{U}(\eta _j\otimes \varphi )|\left(|\xi _j\rangle \langle \xi _i|\otimes d\hat{A}(a)\right)|\hat{U}(\eta _i\otimes \varphi )\rangle \\
 & & = \|\sum _{i=1}^{n} \left(|\xi _i\rangle \langle \xi _i|\otimes d\hat{A}(a)\right)\hat{U}(\eta _i\otimes \varphi )\|^2\\
 & & \ge 0.
 \end{eqnarray*}
Thus every realizable operation measure is completely positive.
The converse statement of this has been proved in
\cite{O84} by
mathematical
construction of the Hilbert space ${\cal K}$ with unit vector $\varphi $,
self-adjoint operator $\hat{A}$ on ${\cal K}$ and unitary operator $\hat{U}$ on
${\cal H}\otimes {\cal K}$ which satisfy Eq.(\ref{341}) for a given 
completely positive operation measure $\cal{I}$ and thus we have
 \begin{Theorem}\label{T341}
 Every completely positive operation measure is realizable.
 \end{Theorem}

For a particular type of measurement statistics, this problem is
the one originally considered by von Neumann. In order to clarify
the relation between the conventional approach and our general
approach, we
shall review his well-known result. Let $X = \sum _i x_i|x_i\rangle \langle x_i|$ be
an observable with a simple discrete spectrum
$\{$\ldots$,x_{-1},x_0,x_1,$\ldots$\}$ and unit eigenvectors $|x_i\rangle $. The
statistics of the precise measurement of $X$ is usually
presupposed as follows: 
 \begin{eqnarray}
 \mbox{measurement probability:} & & P(x_i|\psi ) = |\langle x_i|\psi \rangle |^2,
                                                  \label{3415}\\
 \mbox{state reduction:}         & & \psi \to  \psi [x_i] = |x_i\rangle .
                                                  \label{342}
 \end{eqnarray}
For real numbers $a$ outside the spectrum, we have
$P(a|\psi ) = 0$ and we can put $\psi [a]$ as arbitrary.  This leads to
the operation measure ${\cal I}$ such that
 \begin{equation}
 {\cal I}(\Delta )\hat{\rho} = \sum _{x_i \in \Delta } |x_i\rangle \langle x_i|\hat{\rho}|x_i\rangle \langle x_i|. \label{343}
 \end{equation}
Von Neumann showed that this statistics is consistent with the
second-observer description:  One can construct the probe system
from an arbitrary Hilbert space ${\cal K}$ with basis
$\{$\ldots$,\varphi _{-1},\varphi _{0},\varphi _{1},$\ldots$\}$ .  Let $\varphi _0$ be the probe system
preparation and $\hat{U}$ the unitary operator on ${\cal H}\otimes {\cal K}$
such that
 \[
 \hat{U}(|x_j\rangle \otimes |\varphi _i\rangle ) = |x_j\rangle \otimes |\varphi _{i+j}\rangle ,
 \]
describing the measurement interaction. Let
$A = \sum _i x_i|\varphi _i\rangle \langle \varphi _i|$ be the probe observable. Suppose that the
prior state of the object is $\psi = \sum _kc_k|x_k\rangle $. We have the
following time evolution starting with $\psi \otimes \varphi _0$
 \begin{equation}
 \hat{U}|\psi \otimes \varphi _0\rangle  = \sum _k c_k |x_k\rangle \otimes |\varphi _k\rangle . \label{3435}
 \end{equation}
Then with this 
setting, the joint probability distribution $P(X=x_j,A=x_i)$
obtaining
the result $A = x_i$ and $X = x_j$ in the joint detection just
after the object-probe interaction is
 \begin{eqnarray}
 P(X=x_j,A=x_i)&=& |\langle x_j|\otimes \langle \varphi _i|\hat{U}(\psi \otimes \varphi _0)\rangle |^2 \nonumber\\
               &=& \sum _k|c_k|^2|\langle x_j|x_k\rangle |^2|\langle \varphi _i|\varphi _k\rangle |^2
                                                    \nonumber\\
               &=& |\langle x_j|\psi \rangle |^2\delta _{i,j}.           \label{344}
 \end{eqnarray}
Since, for the second-observer, the measurement probability
$P_{II}(x_j|\psi )$ of this measurement is the probability
$P(A=x_j)$ obtaining the result $A=x_j$, we have from
Eq.(\ref{344}),
 \begin{eqnarray}
 P_{II}(x_j|\psi ) &=& P(A=x_j) = \sum _iP(X=x_i,A=x_j) \nonumber\\
           &=& |\langle x_j|\psi \rangle |^2.                      \label{345}
 \end{eqnarray}
Thus Eq.(\ref{3415}) holds for the second-observer.

Let $P(X=x_j|A=x_i)$ be the conditional probability of obtaining
the result $X=x_j$ given $A=x_i$. Then by Eq.(\ref{344}) we have
 \begin{eqnarray}
 P(X=x_j|A=x_i) &=& \frac{P(X=x_j,A=x_i)}{P(A=x_j)} \\
                &=& \delta _{i,j}.
 \end{eqnarray}
This enables the second observer to make the following
statistical inference: If the first observer were to make the
detection
of the observable $X$ immediately after the fist measurement,
then the results of these two measurement always coincides. This
is possible only if, for the first observer, the first
measurement
changes the state as $\psi \to  |x_j\rangle $ depending on the result $X=x_j$
of the first measurement.  Thus we obtain Eq.(\ref{342})
from the second-observer description. Obviously these reasoning
is a particular case of general consideration presented in
subsection \ref{3.2} and we can thus obtain the operation measure
(\ref{343}) directly from Eqs.(\ref{3.2.9}) and (\ref{3435}) by
the following computation. 
 \begin{eqnarray*}
 {\cal I}(\Delta )(|\psi \rangle \langle \psi |) &=& \int _\Delta d{\cal I}(da)(|\psi \rangle \langle \psi |) \\
 &=& \sum _{x_k\in \Delta }|\psi [x_k]\rangle \langle \psi [x_k]|P(x_k|\psi )\\
 &=& \sum _{x_k\in \Delta }\sum _{i,j} |x_i\rangle 
\langle \hat{U}(\psi \otimes \varphi _0)|\left(|x_j\rangle \langle x_i|\otimes |\varphi _k\rangle \langle \varphi _k|\right)|\hat{U}(\psi \otimes \varphi _0)\rangle \langle x _j|\\
 &=& \sum _{x_k\in \Delta }\sum _{i,j} |x_i\rangle \langle \hat{U}(\psi \otimes \varphi _0)|x_i\rangle |\varphi _k\rangle 
   \langle x_j|\langle \varphi _k|\hat{U}(\psi \otimes \varphi _0)\rangle \langle x _j|\\
 &=&\sum _{x_k\in \Delta }\sum _{i,j} |x_i\rangle  \left(\sum _lc_l^*\langle x_l|x_i\rangle \langle \varphi _l|\varphi _k\rangle \right)
 \left(\sum _lc_l\langle x_j|x_l\rangle \langle \varphi _k|\varphi _l\rangle \right)\langle x_j|\\
 &=&\sum _{x_k\in \Delta }\left(\sum _{i,l}c_l^*|x_i\rangle \langle x_l|x_i\rangle \langle \varphi _l|\varphi _k\rangle \right)
                 \left(\sum _{l,j}c_l  \langle x_j|x_l\rangle \langle \varphi _k|\varphi _l\rangle \langle x_j|\right)\\
 &=&\sum _{x_k\in \Delta } |c_k|^2|x_k\rangle \langle x_k|\\
 &=&\sum _{x_k\in \Delta } |x_k\rangle \langle x_k|\psi \rangle \langle \psi |x_k\rangle \langle x_k|.
 \end{eqnarray*}

Now I shall give some remarks about the conventional approach to
the determination of the state reduction from the second-observer
description.  In the conventional argument, they apply the
so-called projection postulate to the state $\hat{U}(\psi \otimes \varphi )$
(see Eq.(\ref{3435})) just after the interaction
and conclude that if the second observer get the result $A=x_i$
then the state of the object-probe composite system changes into
$|x_i \rangle |\varphi _i\rangle $
and the state of the object changes into
$|x_i\rangle $. Although this argument has an apparent advantage that
it never uses the explicit statistical inference, it has the
several definite weak points.  First, this argument applies only
when the measurement of the probe system is of the first kind and
when it satisfies the projection postulate.  However, any first
kind measurement is subjected to the Schr\"{o}dinger equation for
the object-probe composite system and hence we can never realize
the dynamics causing the projection postulate. Thus
they need to assume a process of the second kind measurement at
some point between the object and the real observer and try to
describe dynamics causing the projection postulate in this
process. However, this means the contradiction that they can use
the projection postulate nowhere since after the second kind
measurement the system state cannot be described by the standard
quantum mechanics and hence the projection postulate cannot apply
to it. Second, their state reduction occurs only after the
detection of the probe system contrary to the fact that the
measurement interaction finishes before the detection of the
probe system. This implies that their argument puts an apparent
limitation for the time interval of the successive measurements.
When one can perform the second measurement of the same system at
the earliest time?  We can say that after the object-probe
interaction but they must say that after the macroscopic
interaction between the probe and the detector. Thus,
from the first point they cannot describe any successive
measurements of the one system.  Last, their argument cannot be
used for the measurement of observables with continuous spectrum
such as the position observable,
since we have no canonical state reduction postulate for
continuous observables. In this case, only statistical inference
can apply. 

Historically speaking, von Neumann did not explicitly use the
projection postulate in the second-observer description. Indeed
he only mentioned the probability correlation between the object
system and the probe system and wrote \lq \lq If III [the second
observer] were to measure (by process 1. [by the subsequent
detection of observables]) the simultaneously measurable
quantities $A$ [the object observable], $B$ [the probe
observable] ( in I [the object system] or II [the probe system]
respectively, or both in I + II ), then the pair of values $a_n$
, $b_n $ would have the probability 0 for $m \ne n$, and the
probability $w_n$ [= the measurement probability for $a_n$] for $m
= n$. $\ldots$ If this is  established, then the measuring process so
far as it occurs in II, is \lq explained' theoretically,$\ldots$''
\cite[p.440]{N55}. (The notes in the brackets above are due to the present 
author.)
 %
 \section{Measurement Breaking SQL}
 \label{4}

 In the preceding sections, we have discussed what is the problem of the SQL
and what we can tell about quantum measurement from quantum mechanics.  One 
conclusion is that every measurement statistics described by a completely
positive operation measure is realizable.  In this last section, I shall 
show that a measurement of the free-mass position which breaks the SQL
is realizable.

   \subsection{Statistics of approximate measurement}
 \label{4.1}

Let ${\cal I}$ be a completely positive operation measure.
Then ${\cal I}$ describes a statistics of a measurement.
Now a problem arise --- when can one consider ${\cal I}$ as a
statistics of a measurement of a given observable?  In what
follows, we shall consider the problem as to when a given operation
measure can be considered as a position measurement of a mass
with one degree of freedom.  Our analysis will lead to
mathematical definitions of precision and resolution of
measurement. This subsection will be concluded with the mathematical
formulation of the SQL.

In the textbook description of position measurement, the
statistics is so characterized as
 \begin{eqnarray}
 \mbox{measurement probability:} & & P(a|\psi ) = |\psi (x)|^2,
                                                  \label{4.1.1}\\
 \mbox{state reduction:}         & & \psi \to  \psi _a = |a\rangle .
                                                  \label{4.1.2}
 \end{eqnarray}
 However, it is proved that there are no operation measures which
satisfies these conditions both. In general, it is proved
\cite{O84} that
there are no weakly repeatable operation measures for
non-discrete observables, where operation measure ${\cal I}$ is called
{\em weakly repeatable} if 
 \begin{equation}
 \mbox{Tr}[{\cal I}(B\cap C)\hat{\rho}] = \mbox{Tr}[{\cal I}(B){\cal I}(C)\hat{\rho}],       \label{4.1.3}
 \end{equation}
 for
all density operators $\hat{\rho}$ and Borel sets $B$, $C$ ---
Eqs.(\ref{4.1.1}) and~(\ref{4.1.2}) lead to this condition. Thus
every
realizable position measurement is an approximate measurement. In
order to clarify the meaning of {\em approximate} measurement, we
shall introduce two error criteria for the preciseness of
measurement.

In Subsection~\ref{2.2}, we have introduced the following
distinction: If the free mass is in a position
eigenstate at the time of a measurement of $x$ then
the {\em precision} $\varepsilon $ of the measurement is defined to
be the uncertainty in the result and the {\em resolution}
$\sigma $ is defined to be the deviation of the position
of the free mass just after the measurement from the
readout just obtained. Now, we shall give
precise definitions for the case of superposition. 

A difficult step in defining the precision is to extract the
noise factor from the readout distribution --- this may be the
reason why the above distinction has hardly ever discussed in
literature.
Let ${\cal I}$ be an operation measure and $\hat{F}$ be its effect measure.
Consider the requirements for ${\cal I} $ to
describe some approximate measurement of
the position observable. Our first requirement is that $\hat{F}$ is
{\em compatible} with the position observable, i.e.,
 \begin{equation}
 {[\hat{x},\hat{F}(\Delta )]} = 0, \quad\mbox{for all Borel sets }\Delta .\label{4.1.4}
 \end{equation}
This condition may be justified by the compatibility of the
information obtained from this measurement with the original
information of the position. Under this condition, there is a
kernel function $G(a,x)$ such that 
 \begin{equation}
 d\hat{F}(a) = da \int  dx\,G(a,x)|x\rangle \langle x|.            \label{4.1.5}
 \end{equation}
 Even if the measuring apparatus measures the position observable
approximately, the readout distribution $P(a|\psi )$ is expected to
be related to the position distribution $|\psi (x)|^2$ --- from
Eq.(\ref{4.1.5}), this relation is expressed in the following
form
 \begin{equation}
     P(a|\psi ) = \int  dx\,G(a,x)|\psi (x)|^2.
                                                  \label{4.1.6}
 \end{equation}
 Note that $G(a,x)$ is independent of a particular wave
function $\psi (x)$ and thus it expresses the noise in the
readout.  Obviously, $P(a|\psi ) = |\psi (a)|^2$ for all $\psi $
(i.e. the noiseless case) if and only if $G(a,x) =
\delta (a - x)$.  From Eq.(9) of \cite{C85}, in the case of von
Neumann's model, $G(a,x) = |\Psi (a - x)|^2$  where $\Psi $
is the prepared state of the probe (See also \cite{C86}, where
$\Psi (a - x)$ is called the resolution amplitude.)
Roughly 
speaking, $G(a,x)$ is the (normalized) conditional
probability density of the readout $a$, given that
the free-mass is in the position $x$ at the time of
measurement; hence the precision $\varepsilon (x)$ of this case
should be 
 \begin{equation}
 \varepsilon (x)^2 = \int  da\,(a - x)^2G(a,x).     \label{4.1.7}
 \end{equation}
 Thus if for the prior state $\psi $ of the mass,
the {\em precision} $\varepsilon (\psi )$ of the
measurement is given by
 \begin{equation}
 \varepsilon (\psi )^2 = \int  dx\,\varepsilon (x)^2|\psi (x)|^2.   \label{4.1.8}
 \end{equation}

By the similar reasoning, for the mass state $\psi $ at the time of
measurement, the {\em resolution} $\sigma (\psi )$ of the measurement is
given by
 \begin{equation}
 \sigma (\psi )^2= \int  da\,\sigma (a)^2P(a|\psi ),     \label{4.1.9}
 \end{equation}
where
 \begin{equation}
 \sigma (a)^2=\int  dx\,(a - x)^2|\psi _a(x)|^2.  \label{4.1.10}
 \end{equation}

The second requirement is that the noise
is {\em unbiased} in the sense that the mean value
of the 
readout is identical with the mean position in the prior state,
i.e.,
 \begin{equation}
 \int  dx\,x P(x|\psi ) = \langle \psi |\hat{x}|\psi \rangle ,      \label{4.1.11}
 \end{equation}
 for all possible $\psi $.  

 Let $\Delta (\psi )$ be the uncertainty of the readout for
the prior state $\psi $ of the mass
and $\Delta x(\varphi )$ the uncertainty of the mass
position at any state $\varphi $.
Then in general we can prove the following
 \begin{Theorem}\label{41T1}
Let ${\cal I} $ be an operation measure for one-dimensional system. Under
conditions of compatibility and unbiasedness expressed by
Eqs.(\ref{4.1.4}) and (\ref{4.1.11}), the following relations hold:
 \begin{eqnarray}
 \varepsilon (\psi )^2 &=& \Delta (\psi )^2- \Delta x(\psi )^2,        \label{4.1.13}\\
 \sigma (\psi )^2 
 &=& \int da\,P(a|\psi )\Delta x(\psi _a)^2+ \int da\,P(a|\psi )(a - \langle \psi _a|\hat{x}|\psi _a\rangle )^2.
                                                  \label{4.1.14}
 \end{eqnarray}
 \end{Theorem}
 Let $\varepsilon _{max}$ be the maximum of $\varepsilon (\psi )$ ranging over
all $\psi $.  Then we have $\varepsilon  _{max} = 0$ if and only if
$P(x|\psi ) = |\psi (x)|^2$ for all $\psi $.

 The above theorem shows that the conditions of compatibility and
unbiasedness are plausible conditions for characterizing
approximate position measurements and for those measurements
further specification concerning approximation can be
done through the precision $\varepsilon $ and the resolution $\sigma $.
Thus in the following, we shall say that an operation measure ${\cal I}$
is of {\em approximate position measurement} if it satisfies the
compatibility condition~(\ref{4.1.4}) and the unbiasedness
condition~(\ref{4.1.11}).

Now we shall give a mathematical formulation of the SQL.
Let ${\cal I} $ be an operation measure for a one-dimensional system.
Suppose that a free mass $m$ undergo unitary evolution during
the time $\tau $ between two identical measurement described by
the operation measure ${\cal I}$.
Let $\hat{U}_\tau $ be the unitary operator of the time evolution.
Suppose that the free mass is in a state $\psi $ just
before the first measurement. Then just after the measurement (at
$t = 0$) the free mass is in the posterior state $\psi _a$
with probability density $P(a|\psi )$. From this readout value $a$,
the observer make a prediction $h(a)$ for the readout
of the second measurement at $t= \tau $.  Then the squared uncertainty
of
this prediction is
 \begin{equation}
 \Delta (\tau ,\psi ,a)^2 = \int  dx\,(x - h(a))^2 P(x|\hat{U}_\tau \psi _a).
                                                  \label{4.1.15}
 \end{equation}
In literature, the following
mean-value-prediction strategy
is adopted for determination of $h(a)$:
 \begin{equation}
 h(a) = \langle \psi _a|\hat{x}(\tau )|\psi _a\rangle ,                        \label{4.1.16}
 \end{equation}
where
 \begin{equation}
 \hat{x}(\tau ) = \hat{U}_\tau ^{\dag}x(0) \hat{U}_\tau  = \hat{x} + \hat{p}\tau /m.       \label{4.4.17}
 \end{equation}
The predictive uncertainty $\Delta (\tau ,\psi )$ of this repeated
measurement with prior state $\psi $ and time duration $\tau $ is defined
as the squared average of $\Delta (\tau ,\psi ,a)$ over all readouts  of the
first measurement,
 \begin{equation}
 \Delta (\tau ,\psi )^2 = \int  da\,\Delta (\tau ,\psi ,a)^2 P(a|\psi ).
                                                  \label{4.1.18}
 \end{equation}
The SQL asserts the relation
 \begin{equation}
 \Delta (\tau ,\psi )^2 \geq  \hbar\tau /m, \quad\mbox{for all prior state }\psi .
                                                  \label{4.1.19}
 \end{equation}
If the operation measure ${\cal I} $ is of approximate position
measurement, we have from Eq.(\ref{4.1.13})
 \begin{eqnarray}
 \Delta (\tau ,\psi ,a)^2 &=& \Delta (\hat{U}_\tau \psi _a)^2,   \nonumber   \\
               &=& \varepsilon (\hat{U}_\tau \psi _a)^2 + \Delta x(\tau )(\psi _a)^2\label{4.1.20}
 \end{eqnarray}
and hence
 \begin{eqnarray}
 \Delta (\tau ,\psi )^2&=&
 \int  da\,P(a|\psi )\left(\varepsilon (\hat{U}_\tau \psi _a)^2 + \Delta x(\tau )(\psi _a)^2\right),
                                                  \nonumber\\
 &=& [\varepsilon (\hat{U}_\tau \psi _a)^2]+[\Delta x(\tau )(\psi _a)^2],         \label{4.1.21}
 \end{eqnarray}
where the brackets means the average due to $P(a|\psi )$.

Now from refinement of Caves's argument in \ref{2.2} (see
Eq.(\ref{2.2.3})), we have the following sufficient condition for
the SQL.

 \begin{Theorem}\label{41T2}
Let ${\cal I} $ be an operation measure of approximate position
measurement with precision $\varepsilon $.
If for any prior state $\psi $ the relation
 \begin{equation}
 [\Delta x(\psi _a)^2] \leq [\varepsilon (\hat{U}_\tau \psi _a)^2]                     \label{4.1.22}
 \end{equation}
holds, then the SQL holds for this measurement, i.e,
 \begin{equation}
 \Delta (\tau ,\psi )^2 \geq  \hbar\tau /m, \quad \mbox{for all prior state
 }\psi.
                                                  \label{4.1.23}
 \end{equation}
 \end{Theorem}

In fact, we have the following estimate using the uncertainty
relation (\ref{2.2.4}),
 \begin{eqnarray*}
 \Delta (\tau ,\psi )^2&=&
 [\varepsilon  (\hat{U}_\tau \psi _a)^2] + [\Delta x(\tau )(\psi _a)^2]\\
 &\geq & [\Delta x(0)(\psi _a)^2] + [\Delta x(\tau )(\psi _a)^2]\\
 &\geq & [2\Delta x(0)(\psi _a)\Delta x(\tau )(\psi _a)] \\
 &\geq & \hbar \tau /m.
 \end{eqnarray*}
The above theorem shows that the concept of precision is relevant
for derivation of the SQL among other candidates for error
of measurement.

If the following simple relation
 \begin{equation}
 \sigma (\psi )^2 \leq [\varepsilon (\hat{U}_\tau \psi _a )^2].  \label{4.1.23b}
 \end{equation}
between the precision and the resolution holds then
assumption~(\ref{4.1.22}) of Theorem~\ref{41T2} follows from
Eq.(\ref{4.1.14}); and hence condition ~(\ref{4.1.23b}) implies the
SQL. Thus it can be said that one's intuition which leads to the SQL
is supported by the following statements; (1) every approximate
measurement satisfies the compatibility  condition and the
unbiasedness conditions for the noise and for the state reduction,
and (2) the resolution is no greater than the precision.  However,
it is not clear at all that every realizable measurement satisfies
the last statement even if the first statement is admitted.

 \subsection{Von Neumann's model of approximate measurement}
 \label{4.2}

In \cite{C85}, Caves showed that von Neumann's model of approximate
position measurement satisfies the assumption of Theorem \ref{41T2}
and so the SQL holds for this model.  Since the object-probe
coupling of von Neumann's model is a simple linear coupling and
it illustrates some proposed models of quantum nondemolition
measurement, we shall review this result in our framework below;
see \cite{N55,C85} for original treatment.

This model of measurement is presented by the second-observer
description. The probe system is a one dimensional system with
coordinate $Q$ and momentum $P$ as well as the object system (the
free mass) with coordinate $x$ and momentum $p$. The object-probe
coupling is turned on from $t=-\tilde{\tau}$ to $t=0$ 
($0 < \tilde{\tau} \ll \tau$ ), it is
described by an interaction Hamiltonian $K\hat{x}\hat{P}$ where $K$ is a
coupling constant and it is assumed to
be so strong that the free Hamiltonians of the mass and the probe
can be neglected; so we choose units such that $K\tilde{\tau} = 1$.  Then if
$\Psi _0(x,Q) = f(x,Q)$, the solution of the Schr\"{o}dinger equation is
 \begin{equation}
 \Psi _t(x,Q) = f(x,Q - Ktx).                        \label{4.1.24}
 \end{equation}
At $t=-\tilde{\tau}$, just before the coupling is turned on, the unknown
free-mass wave function is $\psi (x)$, and the probe is prepared in a
state with wave function $\Phi (Q)$; for simplicity we assume that
$\langle \Phi |\hat{Q}|\Phi \rangle  = \langle \Phi |\hat{P}|\Phi \rangle  = 0$.  The total wave function is
$\Psi _0(x,Q) = \psi (x)\Phi (Q)$.  At the end of the interaction ($t=0$)
the total wave function becomes
 \begin{equation}
 \Psi (x,Q) = \hat{U}\Psi _0(x,Q) = \psi (x)\Phi (Q - x).                 
                                                  \label{4.1.25}
 \end{equation}

In order to obtain the measurement statistics, recall that
the result of this measurement --- the inferred value of $x$ --- is
the value $\overline{Q}$ called the \lq \lq readout'' obtained by the
detection of $Q$ of the probe system turned on at $t=0$ with the
subsequent stage called the \lq \lq detector'' in the overall
macroscopic measuring apparatus. Thus the measurement probability
obtaining the result $\overline{Q}$ is given by the Born statistical
formula
 \begin{equation}
 P(\overline{Q}) = \int dx\,|\Psi (x,\overline{Q})|^2 = \int dx\,|\psi (x)|^2|\Phi (\overline{Q}-x)|^2.
                                                  \label{4.1.26}
 \end{equation}
The posterior wave function $\psi (x|\overline{Q})$ of the free mass (at
$t=0$) is obtained (up to normalization) by evaluating $\psi (x,Q)$ at
$Q = \overline{Q}$:
 \begin{eqnarray}
 \psi (x|\overline{Q}) &=& \frac{\Psi (x,\overline{Q})}{P(\overline{Q})^{1/2}}\nonumber\\
             &=& \frac{\psi (x)\Phi (\overline{Q} - x)}{P(\overline{Q})^{1/2}}
                                                  \label{4.1.27}
 \end{eqnarray}
Note that the posterior wave function~(\ref{4.1.27}) can be
obtained directly from application of Eq.(\ref{3.2.9}) up to
normalization.

To write down the operation measure ${\cal I}$ of this measurement,
notice that if the prior state of the free mass is a mixture $\hat{\rho}$
then the posterior state $\hat{\rho}_{\overline{Q}}$ satisfies the relations
 \begin{equation}
 d{\cal I}(\overline{Q})\hat{\rho} = \hat{\rho}_{\overline{Q}}P(\overline{Q}|\hat{\rho})\,d\overline{Q}
          = \Psi (\overline{Q}1 - \hat{x})\hat{\rho}\Psi (\overline{Q}1 - \hat{x})^{\dag}\,d\overline{Q}.       \label{4.1.28}
 \end{equation}
Thus we have the operation measure ${\cal I}$ of this measurement:
 \begin{equation}
 {\cal I}(\Delta )\hat{\rho} = \int _\Delta   d{\cal I}(\overline{Q})\hat{\rho}
          = \int _\Delta d\overline{Q}\,\Psi (\overline{Q}1 - \hat{x})\hat{\rho}\Psi (\overline{Q}1 - \hat{x})^{\dag}.
                                                  \label{4.1.29}
 \end{equation}
From this the effect measure of this measurement is such that
 \begin{eqnarray}
 \hat{F}(\Delta ) &=& \int _\Delta  d\hat{F}(\overline{Q}),
                                                  \nonumber\\
 d\hat{F}(\overline{Q}) &=& d\overline{Q}\,|\Psi (\overline{Q}1 - \hat{x})|^2
= d\overline{Q}\,\int  dx\,|\Psi (\overline{Q} - x)|^2|x\rangle \langle x|.   \label{4.1.31}
 \end{eqnarray}

From Eq.(\ref{4.1.31}), it is obvious that this measurement
satisfies the
compatibility condition (\ref{4.1.4}) and the kernel function
$G(\overline{Q},x)$ representing the noise in the result can be written as
 \begin{equation}
 G(\overline{Q},x) = |\Psi (\overline{Q} - x)|^2.                          \label{4.1.32}
 \end{equation}
The following computations shows that
the unbiasedness condition~(\ref{4.1.11}) holds:
 \begin{eqnarray}
 \int  d\overline{Q}\,\overline{Q}P(\overline{Q}|\psi )
 &=& \int d\overline{Q}\,dx\,\overline{Q}|\psi (x)|^2|\Phi (\overline{Q}-x)|^2            \nonumber\\
 &=& \int dx\,x|\psi (x)|^2                             \nonumber\\
 &=& \langle \psi |\hat{x}|\psi \rangle .                 \label{4.1.33}
 \end{eqnarray}
Thus this measurement satisfies the conditions for approximate
position measurement.
The precision $\varepsilon (\psi )$ of this measurement is given by
 \begin{eqnarray}
\varepsilon (\psi )^2&=&\int dx\,|\psi (x)|^2\int d\overline{Q}\,(\overline{Q}-x)^2|\Psi (\overline{Q}-x)|^2
                                                  \nonumber\\
        &=& \int d\overline{Q}\,a^2|\Psi (\overline{Q})|^2
                                                  \nonumber\\
        &=& \langle \Psi |\hat{Q}^2|\Psi \rangle .
                                                  \label{4.1.33b}
 \end{eqnarray}
From definition and Eqs.~(\ref{4.1.26})--(\ref{4.1.27}), the
resolution $\sigma (\psi )$ of this measurement is given by
 \begin{eqnarray}
 \sigma (\psi )^2&=& \int d\overline{Q}\,P(\overline{Q}|\psi )\int dx\,(\overline{Q}-x)^2|\psi (x|\overline{Q})|^2
                                                  \nonumber\\
         &=& \int d\overline{Q}dx(\overline{Q}-x)^2|\Psi (\overline{Q}-x)|^2|\psi (x)|^2.
                                                  \nonumber\\
        &=& \langle \Psi |\hat{Q}^2|\Psi \rangle .
                                                  \label{4.1.34}
 \end{eqnarray}
Let $\Delta Q$ be the uncertainty of the probe observable
just before the measurement, i.e., $(\Delta Q)^2 = \langle\Phi|\hat{Q}^2|\Phi
\rangle - \langle\Phi|\hat{Q}|\Phi\rangle^2$. Then in von Neumann's model we have
just obtained
 \begin{eqnarray}
 \varepsilon  &=& \varepsilon (\psi ) = \Delta Q \quad \mbox{for all } \psi ,\\
 \sigma  &=& \sigma (\psi ) = \Delta Q \quad \mbox{for all } \psi .
 \end{eqnarray}
 Thus the assumptions of Theorem~\ref{41T2} hold from the
following computation (cf. Eq.(\ref{4.1.14}))
 \begin{eqnarray*}
 [\varepsilon (\hat{U}_\tau \psi _{\overline{Q}})^2]
 = \Delta Q = \sigma (\psi ) = [\Delta x(\psi _{\overline{Q}})^2] + [(\overline{Q} - \langle \psi _{\overline{Q}}|\hat{x}|\psi _{\overline{Q}}\rangle )^2]
 &\geq & [\Delta x(\psi _{\overline{Q}})^2].
 \end{eqnarray*}
Thus the SQL holds for von Neumann's model of approximate position
measurement.

 \subsection{Realization of measurement breaking SQL}
 \label{4.3}

We shall now turn to Yuen's proposal \cite{Y83}. His observation is
that if the measurement leaves the free mass in a contractive state
$\psi _a$ for every readout $a$ then we can get
 \begin{equation}
 \Delta x(\tau )(\psi _a) \ll (\hbar \tau /2m)^{1/2} \ll \Delta x(0)(\psi _a).
                                                  \label{4.3.1}
 \end{equation}
 Thus the SQL breaks if such a measurement has a good precision
 \begin{equation}
 \varepsilon (\hat{U}_\tau \psi _a) \ll (\hbar \tau /2m)^{1/2}.               \label{4.3.1b}
 \end{equation}
 In fact, from the combination of Eqs.(\ref{4.1.21}) and
(\ref{4.3.1})--(\ref{4.3.1b}), we get
 \begin{equation}
 \Delta (\tau ,\psi )^2 = [\varepsilon (U_\tau \psi _a)^2] + [\Delta x(\tau )(\psi _a)]^2
             \ll  \hbar \tau /m                           \label{4.3.2}
 \end{equation}

In this subsection, I shall show that such statistics of
measurement
can be realized by a measurement considered first by Gordon and
Louisell \cite{G66}.

In \cite{G66}, Gordon and Louisell considered the following
statistical
description of measurement.  Let $\{\Psi _a\}$  and $\{\Phi _a\}$ be a pair
of families of wave functions with real parameter $a$.  The
Gordon-Louisellõ measurement, denoted by $\{|\Psi _a\rangle \langle \Phi _a|\}$, is
the measurement with the following statistics: For any prior
state $\psi $,
 \begin{eqnarray}
 \mbox{measurement probability:} & & P(a|\psi ) = |\langle \Phi _a|\psi \rangle |^2,
                                                  \label{421}\\
 \mbox{state reduction:}         & & \psi \to  \psi _a = \Psi _a.
                                                  \label{422}
 \end{eqnarray}
One of the characteristic properties of the Gordon-Louisell
measurement is that the posterior state $\Psi _a$ depends only
on the measured value $a$ and not at all on the prior state
$\psi $.
For the condition that Eq.(\ref{421}) determines the probability
density, we assume that $\{\Phi _a\}$ is so normalized as
 \begin{equation}
 \int da\,|\Phi _a\rangle \langle \Phi _a| = 1.              \label{423}
 \end{equation}
From Eq.(\ref{423}), it is provided that $\Phi _a$ may not be a
normalizable vector such as position eigenstate $|a\rangle $.  However,
$\Psi _a$ is assumed to be a normalized vector.
The measurement statistics Eqs.(\ref{421})--(\ref{422}) yields the
following operation measure ${\cal I}$ and effect measure $F$:
 \begin{eqnarray}
 {\cal I}(\Delta )\hat{\rho} &=& \int _\Delta  da\, |\Psi _a\rangle \langle \Phi _a|\hat{\rho}|\Phi _a\rangle \langle \Psi _a|, \label{424}\\
  F(\Delta ) &=& \int _\Delta  da\,|\Phi _a\rangle \langle \Phi _a|.                  \label{425}
 \end{eqnarray}

The following computation shows that the operation measure ${\cal I}$
in
Eq.(\ref{424}) is completely positive: For any finite sequences
of vectors $\xi _1,\xi _2,$\ldots$,\xi _n$ and $\eta _1,\eta _2,$\ldots$,\eta _n$, we obtain
 \begin{eqnarray*}
 \sum _{i,j=1}^{n} \langle \xi _i|d{\cal I}(a)(|\eta _i\rangle \langle \eta _j|)|\xi _j\rangle  &=&
 da\, \sum _{i,j=1}^{n} \langle \xi _i|\Psi _a\rangle \langle \Phi _a|\eta _i\rangle \langle \eta _j|\Phi _a\rangle \langle \Psi _a|\xi _j\rangle  \\
 &=& da\,\left|\sum _{i=1}^{n} \langle \xi _i|\Psi _a\rangle \langle \Phi _a|\eta _i\rangle \right|^2 \\
 &\geq & 0.
 \end{eqnarray*}
 Thus from Theorem~\ref{T341}, we get the following

 \begin{Theorem}
 Every Gordon-Louisell measurement is realizable.
 \label{T.4.3.1}\end{Theorem}

As mentioned above, Gordon-Louisell measurements controls the
posterior states independently of the prior states and this
properties are suitable for our purpose of realization
of measurement which leaves the free mass in a contractive state.
The following Theorem is an immediate consequence from
Theorem~\ref{T.4.3.1}, which asserts that the state reduction of
position measurement can be arbitrarily controlled.

 \begin{Theorem}
 For any Borel family $\{\Psi _a\}$ of unit vectors,
the following statistics of position measurement is realizable:
 \begin{eqnarray}
 \mbox{\rm measurement statistics:}& & P(a|\psi ) = |\psi (a)|^2,\\
 \mbox{\rm state reduction:}      & & \psi \to  \psi _a =\Psi _a.
 \end{eqnarray}
This measurement corresponds to the Gordon-Louisell measurement
$\{|\Psi _a\rangle \langle a|\}$ with the following operation measure
${\cal I} $
and effect measure  $\hat{F}$:
 \begin{eqnarray}
 {\cal I} (\Delta )\hat{\rho} &=& \int _\Delta  da\, |\Psi _a\rangle \langle a|\hat{\rho}|a\rangle \langle \Psi _a|,
                                          \label{4.3.8}\\
 \hat{F}(\Delta ) &=& \int _\Delta  da\,|a\rangle \langle a|. \label{4.3.9}
 \end{eqnarray}
 \label{T.4.3.2}\end{Theorem}

 Let $|\mu \nu 0 \omega \rangle $ be a fixed contractive state with $\langle x\rangle  = \langle p\rangle 
= 0$ and let $\Psi _a$ be such that $\langle x|\Psi _a\rangle  = \langle x - a|\mu \nu  0 \omega \rangle $.
Then $\{\Psi _a\}$ is the family of contractive states $|\mu \nu a \omega \rangle $
with $\langle x\rangle  = a$ and $\langle p\rangle =0$, which satisfies the assumption in
Theorem~\ref{T.4.3.2}. Thus the following measurement
statistics of Gordon-Louisell measurement $\{|\mu \nu a \omega \rangle \langle a|\} $ is 
realizable:
 \begin{eqnarray}
 \mbox{measurement probability:}& & P(a|\psi ) = |\langle a|\psi \rangle |^2,
                                             \label{4.3.10}\\
 \mbox{state reduction:}        & & \psi \to  \psi _a = |\mu \nu a \omega \rangle .
                                             \label{4.3.11}
 \end{eqnarray}

Now we shall examine the predictive uncertainty $\Delta (\tau )$ of the
repeated measurements of this measurement.
From Eqs.(\ref{4.3.9}) and (\ref{4.3.10}), this measurement
satisfies the compatibility condition (\ref{4.1.4}) and the
unbiasedness condition (\ref{4.1.11}). Further, it is an exact
measurement of the position observable, i.e., $G(a,x) = \delta (a - x)$
and $\varepsilon (\psi ) = 0$ for all $\psi $.
From Eqs.(\ref{4.1.21}) and (\ref{2.1.16})--(\ref{2.1.17}), we have
 \begin{eqnarray*}
 \Delta (t,\psi )^2 &=&
 \int  da\,P(a|\psi )\left(\varepsilon (\hat{U}_t \psi _a)^2 + \Delta x(t)(\psi _a)^2\right),\\
 &=& \int  da\,P(a|\psi ) \Delta x(t)(\psi _a)^2,\\
 &=&(1/4\xi )(\hbar \tau /m) + (2\hbar /m\omega )(|\mu +\nu |\omega /2)^2(t-\tau )^2,
 \end{eqnarray*}
 where
 \[
 \tau  =  2\xi /(\omega |\mu +\nu |^2) = \xi \hbar m/\Delta p(0)^2.
 \]
Thus for $t=\tau $ and large $\xi $, we have
 \[
 \Delta (\tau ,\psi )^2 = (1/4\xi )(\hbar \tau /m) \ll \hbar \tau /m.
 \]

We have therefore shown that the Gordon-Louisell measurement ${|\mu \nu a \omega \rangle \langle a|}$
is realizable measurement and it breaks the SQL. In the rest of
this subsection, I shall give a realization of this measurement
with an
interaction Hamiltonian of the object-probe coupling \cite{O88}.

The model description is parallel with
that of von Neumann's measurement in \ref{4.2}.
The free mass (the object system) is coupled to a probe system
which is
a one dimensional system with coordinate $Q$ and
momentum $P$.  The coupling is turned on from $t =
-\tilde{\tau}$ to $t=0$ ($0 < \tilde{\tau} \ll \tau $) and it is assumed
to be so strong that the free Hamiltonians of the
mass and the probe can be neglected.  We choose the
following interaction Hamiltonian
 \begin{equation}
 H = \frac{K\pi }{3\sqrt 3}\{2(\hat{x}\hat{P} - \hat{Q}\hat{p})+(\hat{x}\hat{p}-\hat{Q}\hat{P})\},
 \end{equation}
where $K$ is the coupling constant chosen
as $K\tilde{\tau}=1$. Then if $\Psi _0(x,Q) = f(x,Q)$, the
solution of the Schr\"{o}dinger equation is
 \begin{eqnarray}
 \lefteqn{\Psi _t(x,Q) = f\left(\frac{2}{\sqrt 3}
 \left\{x\sin\frac{\left(1-Kt\right)\pi }{3}
 +Q\sin\frac{Kt\pi }{3}\right\}\right.,} \nonumber\\
 & & \hfill\left.\frac{2}{\sqrt 3}\left\{-x\sin\frac{Kt\pi }{3}
 +Q\sin\frac{\left(1+Kt\right)\pi }{3}\right\}\right).
 \end{eqnarray}
At $t=-\tilde{\tau}$, just before the coupling is turned on,
the unknown free-mass wave function is $\psi (x)$, and
the probe is prepared in a contractive state $\Phi (Q) =
\langle Q|\mu \nu 0\omega \rangle $, so that the total wave function is
$\Psi _0(x,Q) = \psi (x)\Phi (Q)$; expectation values for this
state is $\langle \hat{Q}\rangle _0=\langle \hat{P}\rangle _0=0$. At $t=0$, the end of the
interaction, the total wave function becomes 
 \begin{equation}
 \Psi (x,Q) = \psi (Q)\Phi (Q-x).
 \end{equation}
Compare with Eq.(\ref{4.1.25}); the statistics is
much different.  At this time, a value
$\overline{Q} $ for $Q$ is obtained by the detector of the probe
observable in the subsequent stage of the measuring apparatus,
from which one infers a value for $x$. Thus the
probability density $P(\overline{Q} |\psi )$ to obtain the
value $\overline{Q} $ as the result of this measurement is
given by
 \begin{equation}
 P(\overline{Q} |\psi ) = \int dx |\Psi (x,\overline{Q} )|^2 = |\psi (\overline{Q}
)|^2.
 \end{equation}
The free-mass wave function $\psi _{\overline{Q}} (x)=\psi (x|\overline{Q}
)$ just after this measurement ($t=0$) is
obtained (up to normalization) by
 \begin{eqnarray*}
 \psi (x|\overline{Q} ) &=&  [1/P(\overline{Q} |\psi )^{1/2}]\Psi (x,\overline{Q} )
\\
             &=& [\psi (\overline{Q} )/|\psi (\overline{Q} )|]\Phi (\overline{Q} -
x) \\
             &=& C\langle x|\mu \nu \overline{Q}\omega \rangle ,
 \end{eqnarray*}
where $C$ ($|C|=1$) is a constant phase factor.

Thus we have just obtained the measurement statistics of this
measurement as follows:
 \begin{eqnarray}
 \mbox{measurement probability:} & & P(\overline{Q}|\psi ) = |\psi (\overline{Q} )|^2,\\
 \mbox{state reduction:} & & \psi  \to  \psi _{\overline{Q}} = |\mu \nu \overline{Q} \omega \rangle .
 \end{eqnarray}
This shows that this measurement is a realization of
Gordon-Louisell measurement $\{|\mu \nu \overline{Q}\omega \rangle \langle \overline{Q}|\}$.

Now I have shown everything I promised before. From our analysis,
we can conclude that there are no general reasons in physics
which limits the accuracy of the repeated measurement of the
free-mass position such as the standard quantum limit for
monitoring the free-mass position.  In \cite{B86}, Bondurant
analyzed the performance of a interferometric gravity-wave detector
which has
a Kerr cell in each arm used to counter the effects of radiation
pressure fluctuation and has a feedback loop used to keep the
interferometer operating at the proper null.  He succeeded in
showing that this measurement realizes a monitoring the free-mass
position which breaks the SQL.
It will be an interesting problem to show that the role of the Kerr
cell and the feedback loop in his analysis has some corresponding
part in the interaction scheme realizing the Gordon-Louisell
measurement discussed above.
 
  \end{document}